\documentclass[aps,prl,showpacs,amsmath,amssymb,twocolumn,nofootinbib]{revtex4}
\usepackage{pslatex,graphicx,dcolumn,bm,natbib}
\date{\today}
\usepackage{epsfig}
\usepackage{color}

\begin{document}

\title{Jamming in Systems Composed of Frictionless Ellipse-Shaped
Particles}

\author{Mitch Mailman$^{1}$, Carl F. Schreck$^{2}$, Corey S. O'Hern$^{2}$ and Bulbul Chakraborty$^{1}$\\
\normalsize{$^{1}$Martin Fisher School of Physics, Brandeis University, Mail Stop 057,
Waltham, MA 02454-9110, USA\\
$^{2}$Department of Mechanical Engineering, Yale University, New Haven,
CT 06520-8284 and Department of Physics, Yale University, New Haven
CT 06520-8120}}

\begin{abstract}
We study the structural and mechanical properties of jammed ellipse
packings, and find that the nature of the jamming transition in these
systems is fundamentally different from that for spherical particles.
Ellipse packings are generically hypostatic with more degrees of
freedom than constraints.  The spectra of low energy excitations
possess two gaps and three distinct branches over a range of aspect
ratios. In the zero compression limit, the energy of the modes in the
lowest branch increases {\it quartically} with deformation amplitude,
and the density of states possesses a $\delta$-function at zero
frequency. We identify scaling relations that collapse the
low-frequency part of the spectra for different aspect
ratios. Finally, we find that the degree of hypostaticity is
determined by the number of quartic modes of the packing.
\end{abstract}
\maketitle 

A decade ago, Liu and Nagel proposed that jamming transitions in
glassy, granular, and other nonequilibrium amorphous systems can be
described by the jamming phase diagram \cite{LiuNagel}, and that a
`fixed point' (Point J) in the jamming phase diagram controls slow
dynamics in these systems even far from Point J \cite{OHern2003}.  For
model disordered systems composed of frictionless, spherical grains
interacting via purely repulsive, short-range potentials, the packings
at Point J are \emph{isostatic} \cite{Makse2000,OHern2002}, where the
number of degrees of freedom exactly matches the number of
constraints.  It has been shown that isostatic packings of spherical
grains have interesting mechanical properties; for
example, they possess an abundance of spatially extended ``floppy
modes'' of excitation \cite{Wyart1,Wyart2,somfai} and non-elastic
response to applied deformations \cite{moukarzel}.  

However, there have been relatively few theoretical studies of jamming
in systems with aspherical particles, despite the fact that most
physical particulate media have grains with anisotropic shapes.  The
introduction of aspherical particles in equilibrium systems gives rise
to completely new phases of matter and critical behavior as evidenced
in the field of liquid crystals.  What is the nature of jamming
transitions in nonequilibrium systems when the grains are
aspherical?  For example, is there still a special point J in the
jamming phase diagram, where packings are isostatic, that controls
slow dynamics?  In this letter, we begin to address these important open
questions by investigating the structural and mechanical properties of
static packings of frictionless anisotropic particles.  

For a static packing of $N$ grains in $d$ spatial dimensions, with
$d-1$ rotational and $d$ translational degrees of freedom per grain,
total force and torque balance on each grain can be satisfied only if
the total number of contacts satisfies $N_c \ge N_I \equiv
N\left(2d-1\right)$ in the large system limit.  Isostatic packings
satisfy $N_c = N_I$, while hypostatic packings possess $N_c < N_I$.
In contrast to spherical particle packings, static packings of
ellipsoidal particles, studied previously by Donev, {\it et al.}
\cite{donev1}, are generically hypostatic and possess higher densities
without translational and orientational order.  Experiments on
packings of ellipsoidal $M\&M$ candies have also verified these
results\cite{donev2}.  These previous studies raise several
fundamental questions about static packings of ellipsoidal particles;
for example, why are they hypostatic and what is the nature of their
low-energy excitations?  In this letter, we study the low-energy,
vibrational excitations of 2D static packings of ellipses using a
numerical packing-generation algorithm in which soft ellipses interact
via purely repulsive potentials at zero temperature.  Our analysis
demonstrates the existence of two gaps in the vibrational spectrum
over a range of aspect ratios. The energy of the modes below the first
gap increases {\it quartically} with deformation amplitude along the
soft directions, and the number of these quartic modes determines the
degree of hypostaticity of the packings.

\paragraph{Compression packing-generation protocol} 
We generated an extensive set of static packings of ellipses over a
range of system sizes from $N=120$ to $1920$, in which particles are
`just touching', using a numerical packing-generation protocol similar
to that employed to create static packings of spherical particles
\cite{Corey_method,makse_method}.  We will refer to this protocol as
the `compression method'. In this method, soft, purely repulsive
ellipses are first randomly deposited in a square cell with
periodic boundary conditions at a low packing fraction ($\phi=0.5$).
The configurations are successively compressed in small steps $(\Delta
\phi = 10^{-4})$ and then relaxed using conjugate energy minimization
after each step.  Near the jamming packing fraction $\phi_J$, where
the particles are just touching, the configurations are expanded or
compressed by decreasing amounts until the system has vanishingly
small total potential energy per particle $V_{\rm tol} < V < 2 V_{\rm
tol}$.  $V_{\rm tol} = 10^{-12}$ for most simulations, $V = \sum_{i>j}
V(r_{ij})$ summed over all ellipse pairs,
\begin{equation}
V(r_{ij})=\begin{cases}
\left(1-r_{ij}/\sigma_{ij}\right)^{2} &  r_{ij} \leq \sigma_{ij}\\
0 & r_{ij} > \sigma_{ij},
\end{cases}
\label{eq:pair_pot}
\end{equation}
and $r_{ij}$ is the separation between the centers of mass of ellipses
$i$ and $j$.  The separations and orientations ${\hat u}_i$
and ${\hat u}_j$ of the long axes of ellipses $i$ and $j$ determine
the Perram and Wertheim overlap parameter
\cite{Geo_Overlap_Der,Geo_Overlap_Der_More,BP,Cleaver_Bidisperse}
\begin{eqnarray}
\sigma_{ij} &=& \min_\lambda
\frac{\sigma_{0}(\lambda)}{\sqrt{1-\frac{\chi(\lambda)}{2} \sum_{\pm}
\frac{\beta(\lambda)\hat{r}_{ij}\cdot\hat{u}_i\pm\beta^{-1}(\lambda)
\hat{r}_{ij}\cdot\hat{u}_j}
{1\pm\chi(\lambda)\hat{u}_i\cdot\hat{u}_j}} } ,
\label{eq:sigma2}
\end{eqnarray} 
where $\sigma_0$, $\beta$, and $\chi$ depend on $\lambda$ and the
major (minor) axis, $a_i$ ($b_i$), of the $i$th ellipse \cite{foot}.
To determine $\sigma_{ij}$ for each pair, minimization of the
parameter $0 < \lambda < 1$ must be performed.  We simulate bidisperse
mixtures to inhibit translational and orientational order: one-third
of the particles are large with the major axis $1.4$ times that of the
small particles~\cite{donev1}.  Using this procedure, we generated an
ensemble of at least $100$ ellipse packings, each characterized by the
jamming packing fraction $\phi_J$, over a range of aspect ratios from
$\alpha=1$ to $2$. We calculated the global bond-orientational and
nematic order parameters \cite{frenkel}, and found no significant
ordering with order parameters $\sim 1/\sqrt{N}$ for all $\alpha$.

\paragraph{Vibrational Spectra}
\begin{figure}
\epsfig{file=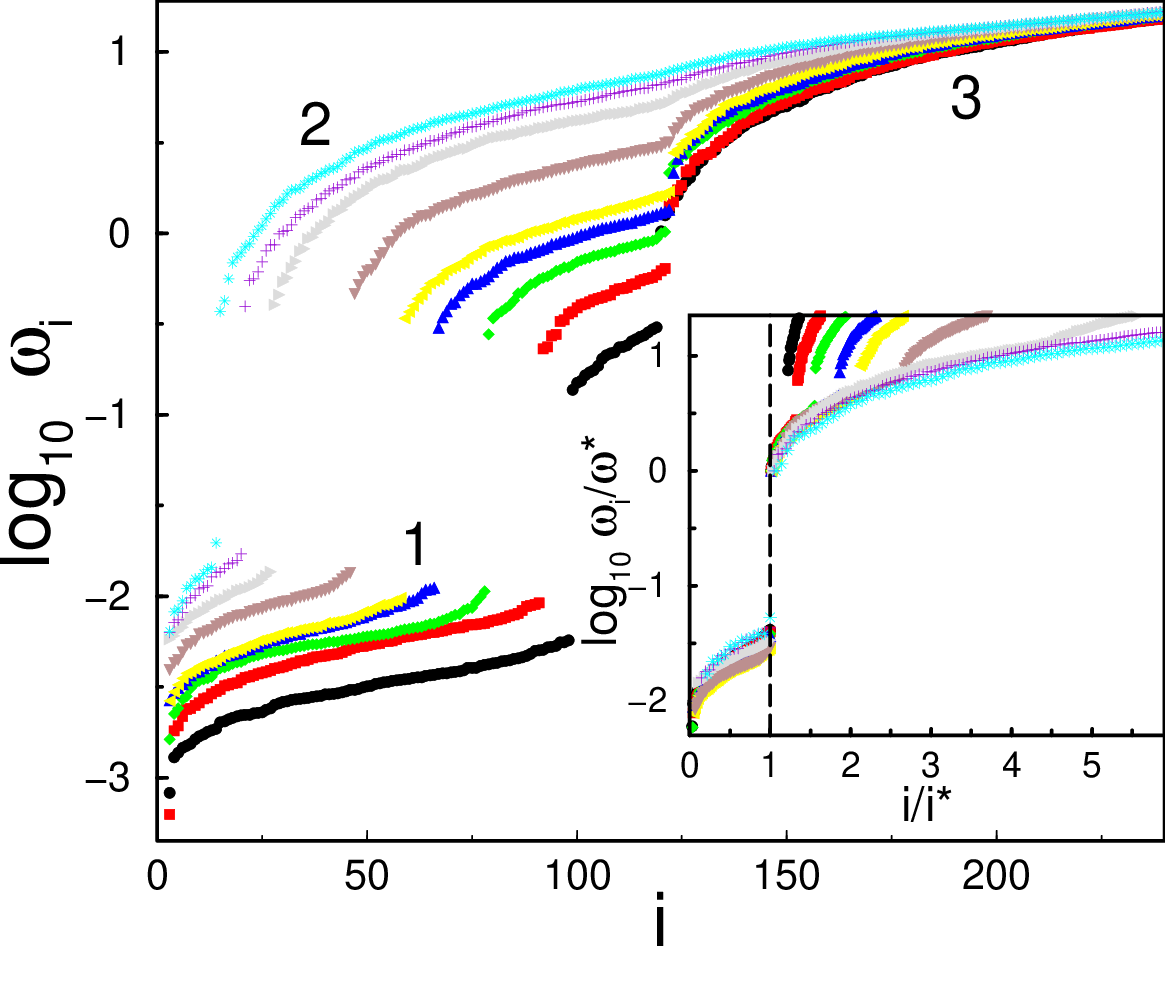, angle=-90, width=1.0\linewidth}
\vspace{-0.2in}
\caption{(Color online) Normal mode frequencies $\omega_i$ from the
dynamical matrix (Eq.~\ref{dyn_matrix}) vs. index $i$, sorted by
increasing frequency for $N=120$ ellipse packings at nine
aspect ratios, $\alpha=1.02$ (black), $1.04$ (red), $1.06$ (green),
$1.08$ (blue), $1.1$ (yellow), $1.2$ (brown), $1.4$ (gray), $1.6$
(violet), and $1.80$ (turquoise).  The sorted frequency spectrum
possesses three distinct branches (numbered $1$, $2$, and $3$).  In
the inset, we show the scaled frequency $\omega_i/\omega^*$
vs. $i/i^*$, which collapses the low frequency part of the spectra at
1 (vertical dashed line).
\label{fig:plot1}}
\vspace{-0.2in}
\end{figure}

To determine mechanical properties of ellipse packings, we calculate
the dynamical matrix 
\begin{equation}
M_{mn}=\frac{\partial^2
V} {\partial \xi_{m}\partial \xi_{n}},
\label{dyn_matrix}
\end{equation}
where $\xi_m=\{x_m,y_m,a_m \theta_m\}$, $x_m$
and $y_m$ are center of mass coordinates of ellipse
$m$, $\theta_m$ is the angle between ${\hat u}_m$ and the $x$-axis,
and $m,n=1,\cdots,N$ \cite{barrat}.  When Eq.~\ref{dyn_matrix} is
evaluated for an ellipse packing and diagonalized using periodic
boundary conditions, in principle one obtains $(2d-1)N' -d$ nontrivial
vibrational eigenmodes, where $N' = N - N_r$ and $N_r$ is the number
of `rattlers' with fewer than $d+1$ contacts.  If we assume that all
ellipses have the same mass, the square roots of the eigenvalues of the
dynamical matrix give the normal mode frequencies $\omega_i$ indexed
by $i$.  We denote the normalized eigenvector corresponding to
$\omega_i$ by ${\hat e}_i = \{e_{xi}^{j=1},e_{yi}^{j=1},e_{\theta
i}^{j=1},\cdots,e_{xi}^{j=N'},e_{yi}^{j=N'},e_{\theta i}^{j=N'}\}$
with the constraint that ${\hat e}^2_i = 1$.  Below, we show the
relative contributions of translational and orientational components
of the eigenvectors, for example, the translational contribution from
mode $i$ is $T_i = \sum_{j=1}^{N'}\lbrace (e_{xi}^j)^2 + (e_{yi}^j)^2
\rbrace$ and $R_i = 1 - T_i$.

\begin{figure}
\epsfig{file=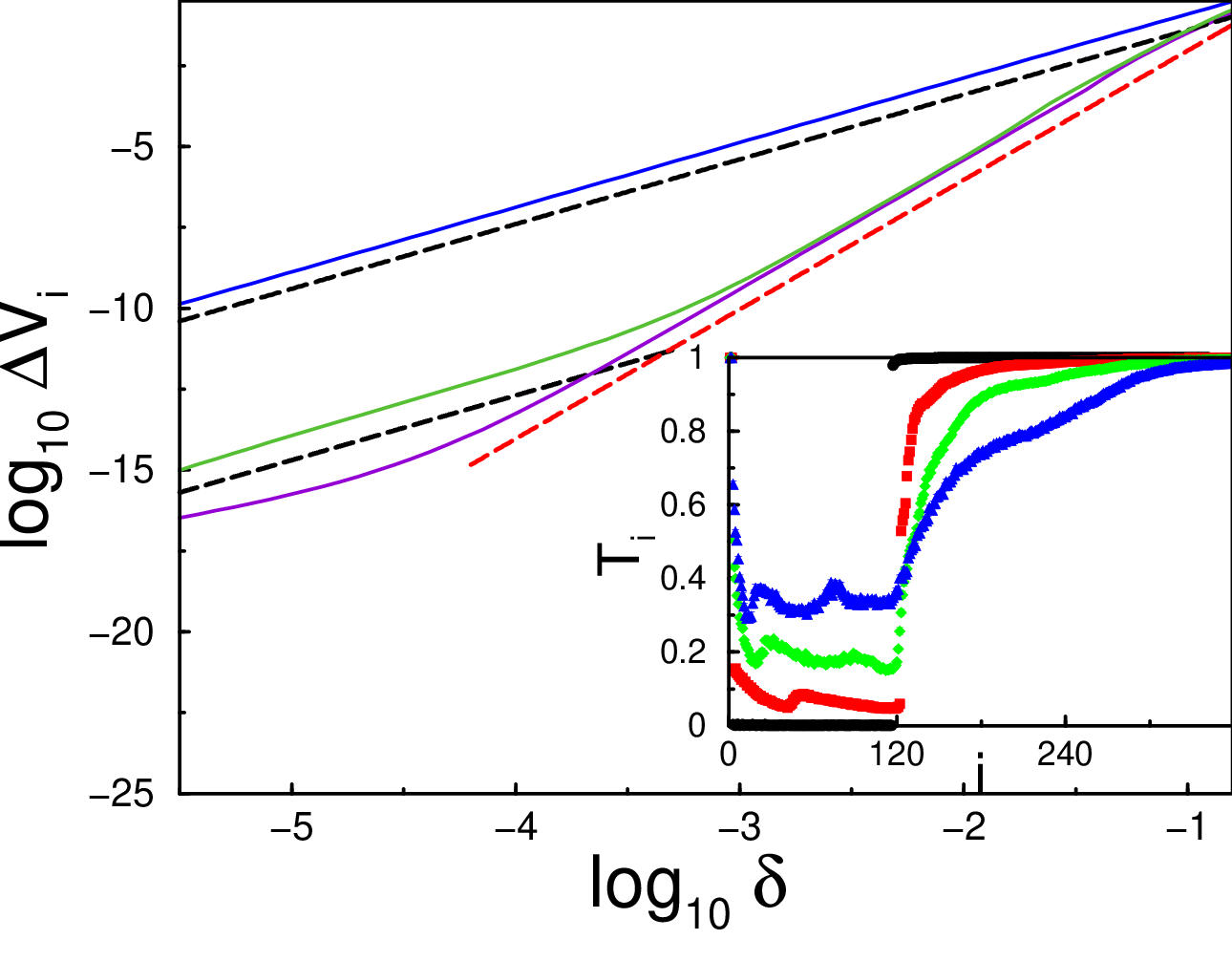, angle=-90, width=1.0\linewidth}
\vspace{-0.15in}
\caption{(Color online) Change in potential energy $\Delta V_i$ vs. 
displacement $\delta$ along ${\hat e}_i$ for $N=120$
and $\alpha=1.5$ ($i^{*}=22$).  $\Delta V_i$ (solid blue) for $i=115 > i^*$ is
quadratic in $\delta$.  In contrast,
for $i\leq i^*$, $\Delta V_i \propto \delta^{2}$ for
$\delta<\delta_c$, but $\propto \delta^{4}$ for $\delta>\delta_c$.
For $i=24$, we show that $\delta_c$ decreases from approximately
$10^{-3}$ to $10^{-4}$ as $V_{\rm tol}$ varies from $10^{-12}$ (green)
to $10^{-16}$ (purple).  The dashed black (red) line has slope two (four).
Inset: The translational contribution $T_i$
to the sum of the squares of the amplitudes of each eigenvector ${\hat
e}_i$ of the dynamical matrix for aspect ratio $\alpha=1.01$ (black),
$1.20$ (red), $1.50$ (green), and $2.00$ (blue).}
\label{fig:plot2}
\vspace{-0.2in}
\end{figure}

Over a range of aspect ratios, the spectrum $\omega_i$, sorted in
order of increasing frequency, possesses three distinct regimes ({\it
cf} Fig.~\ref{fig:plot1}): (1) modes with indexes $i - 2 <
i^*(\alpha)$ below the low-frequency gap, (2) modes with index $i^*
\leq i-2 \leq i_t = (d-1)N$, where for $\alpha \le \alpha_t$, there is
a second gap at index $i_t$, and (3) modes with index $i-2 > i_t$.
(We are explicitly not including the two trivial modes corresponding
to translational invariance.)  In the inset to Fig.~\ref{fig:plot1},
we show that we are able to choose aspect ratio dependent scaling
factors $\omega^*$ and $i^*$ that collapse the low-frequency part of
the spectra including the first gap.  We find that $\omega^*$ is
roughly linear with $\alpha-1$, while $i^*$ possesses two different
scaling regimes for $\alpha-1 \ll 1$ and $\alpha > 1$.  As
demonstrated in the inset to Fig.~\ref{fig:plot2}, modes in regions
(1) and (2) are mainly rotational, whereas those in region (3)
correspond to mainly translational motion.

\begin{figure}
\epsfig{file=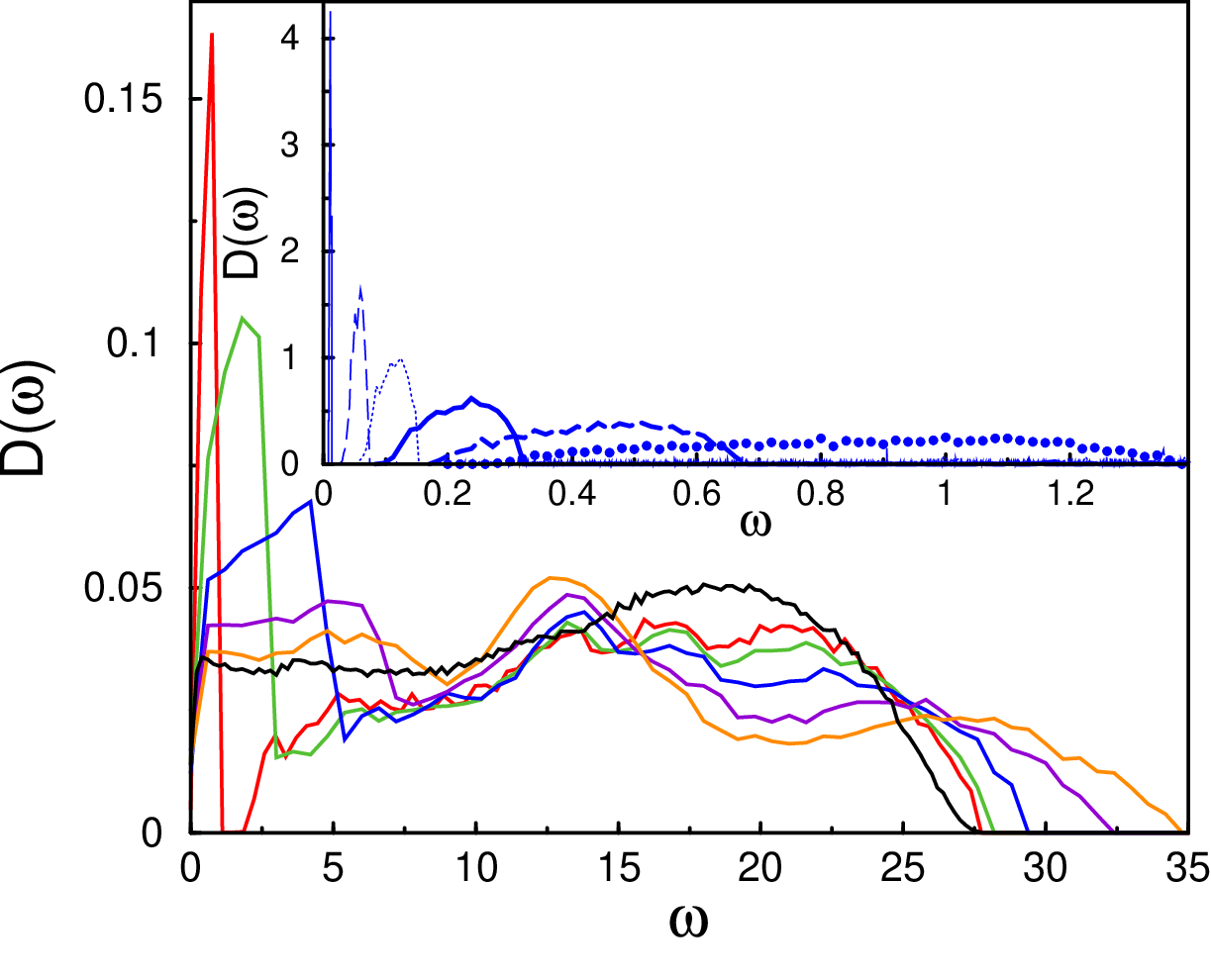, angle=-90, width=0.95\linewidth}
\vspace{-0.2in}
\caption{Density of vibrational modes $D(\omega)$ for ellipse packings
with $N=120$ for $\alpha=1.0$ (black), $1.1$ (red), $1.2$ (green),
$1.4$ (blue), $1.7$ (violet), and $2.0$ (orange). The low $\omega$
peak corresponds to rotational modes in frequency regime $2$ discussed
in the text. The inset magnifies the low $\omega$ region for $\alpha$
near $1$: $\alpha=1.001$ (thin solid), $1.005$ (thin dashed), $1.01$
(thin dotted), $1.02$ (thick solid), $1.04$ (thick dashed), and $1.08$
(filled circles).  The low-frequency peak is sharp and $D(\omega)$
possesses a gap in $\omega$ for $\alpha < \alpha_t \approx 1.2$.
However, the peak broadens and connects to $D(\omega)$ at large
$\omega$ without going to zero for $\alpha>\alpha_t.$}
\label{fig:DOS}
\vspace{-0.2in}
\end{figure}

We find that our ellipse packings at finite, but small overcompression
possess $(2d-1)N-d$ {\it nonzero, positive} eigenvalues \cite{foot2}.
To rationalize this result with the fact that hard ellipse packings
are generically hypostatic~\cite{donev1}, we perturbed our ellipse
packings along each of the eigendirections of the dynamical matrix
over a range of overcompression.  If $\vec{\xi}_0$ characterizes the
centers of mass and orientations of the original static ellipse
packing, the perturbed configuration obtained after a shift by
$\delta$ along eigenmode $i$ and relaxation to the nearest local
energy minimum is $\vec{\xi}_i = \vec{\xi}_{0} + \delta {\hat e}_i$.
In Fig.~\ref{fig:plot2}, we plot the change in potential energy,
$\Delta V_i \equiv V(\vec{\xi}_i)-V(\vec{\xi}_{0})$, of ellipse
packings with $N=120$ at $\alpha=1.5$ arising from a perturbation
along mode $i$ as a function of amplitude $\delta$ and for two values
of overcompression $V_{\rm tol}$.  As shown in Fig.~\ref{fig:plot2},
for modes with indexes in regions (2) and (3) of the frequency
spectrum $\Delta V_i \propto \delta^{2}$ for all $\delta$ independent
of $V_{\rm tol}$.  In contrast, there is range $\delta > \delta_c$
over which modes in region (1) display {\it quartic} dependence on
$\delta$, $\Delta V_i \propto \delta^{4}$, whereas $\Delta V_i \propto
\delta^{2}$ for $\delta < \delta_c$.  Since $\delta_c \sim V_{\rm
tol}^{1/4}$ for modes in region (1), quartic behavior will persist
over the entire range of $\delta$ in the zero compression limit.
Thus, `just-touching' ellipse packings are stabilized by
quartic rather than quadratic terms in the expansion of the total
potential energy around the reference packing\cite{donev1}.

The density of vibrational modes, $D(\omega)$, obtained from the
spectrum, $\omega_{i}$, shown in Fig.~\ref{fig:DOS}, exhibits several
key differences from that for disk packings \cite{OHern2003}.  First,
the plateau at small $\omega$ characteristic of nearly-isostatic disk
packings is replaced by a narrow peak at low frequencies.  This peak
is composed of modes in region (2) of the spectrum that display
collective, primarily rotational motions.  For $\alpha < \alpha_t$
(with $\alpha_t \approx 1.2$ for $N=120$), this peak is clearly
separated by a gap from the broad, high-frequency regime.  Note that
on a logarithmic scale one would also see a peak that corresponds to
modes in region (1), and is separated by a gap from the modes in
region (2).  As discussed before, this peak shifts to lower $\omega$,
and narrows to a $\delta$-function at zero frequency in the
zero-compression limit. Second, for large aspect ratios ($\alpha >
1.5$ for $N=120$), $D(\omega)$ has significantly more structure than
that for disk packings for $\omega > 10$.  These new features will be
investigated in more detail in future work \cite{future}.  We have
verified that the vibrational properties described here also hold for
packings in which ellipses interact via the Gay-Berne potential
\cite{GB} with a simpler form for the overlap function than in
Eq.~\ref{eq:sigma2}.
 
\paragraph{Hypostaticity}
\begin{figure}
\epsfig{file=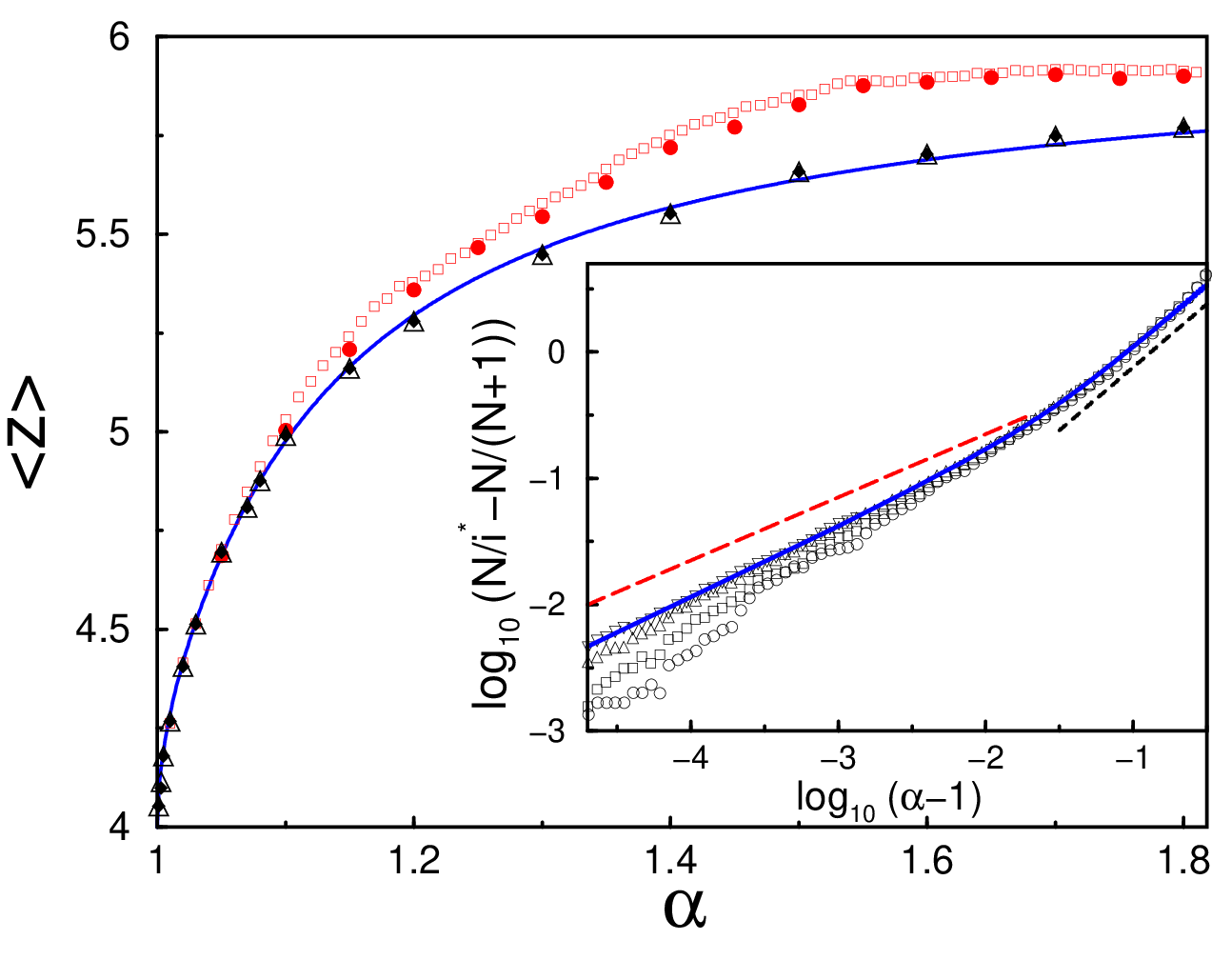, angle=-90, width=1.0\linewidth}
\vspace{-0.2in}
\caption{(Color online) Contact number $\langle z \rangle$
vs. $\alpha$ for compression (triangles) and annealing
($\Delta\alpha=0.005$ (open squares) and $0.05$ (filled circles))
methods for $N=480$.  $\langle z \rangle$ for annealing diverges from
that for compression for $\alpha > 1.1$. The filled diamonds represent
$\langle z \rangle = 6 - i^*(\alpha)/(3N)$ for the compression method
data.  The inset shows the system size dependence of $N/i^*(\alpha) -
N/(N+1)$ for $N=120$ (circles), $240$ (squares), $960$ (upward
triangles), and $1920$ (downward triangles) for the compression data.
The red (black) dashed line has slope $0.5$ ($1.0$).  The solid line
is an interpolation between power-laws of $0.5$ at small $\alpha-1$
and $1.0$ at large $\alpha-1$, which was used to fit $\langle
z\rangle$ in the main plot (solid blue line).}
\label{fig:plot4}
\vspace{-0.2in}
\end{figure}

Our analysis of the structural and mechanical properties of ellipse
packings has led to several novel observations: (1) The quartic modes
that exist for hard ellipse packings develop a quadratic contribution
at finite overcompression, and therefore compression {\it stabilizes}
ellipse packings\cite{donev1}; (2) The contribution of these quartic
modes to the density of vibrational modes decreases as the aspect
ratio increases; and (3) Ellipse packings are generically hypostatic,
as shown in Fig.~\ref{fig:plot4}, with smaller average contact number
$\langle z \rangle$ than that predicted from isostatic counting
arguments, i.e. $\langle z \rangle_{\rm iso} = 6$ for ellipses with
$\alpha>1$.  Hypostatic packings have fewer contacts than required to
satisfy all force and torque balance conditions, and thus in
some directions of configuration space these packings are only
quartically, not quadratically stable.  We find that quartic modes
represent collective, primarily rotational motions of the ellipses,
which do not lead to cage breaking and particle rearrangements.  Thus,
we expect that if the isostatic counting argument is reformulated so
that the quartic modes are not constrained, $\langle z \rangle$
will correspond to the minimum number of contacts necessary to
constrain the quadratic modes.  Thus, ellipse packings are isostatic
with respect to only the {\it quadratic} modes.

When each internal degree of freedom in an ellipse packing is
stabilized, the isostatic conjecture gives $N \langle z
\rangle/2=3N-1$.  If it is not necessary to constrain the quartic
modes, this equation can be rewritten as
\begin{equation}
\frac{N \langle z \rangle}{2}= 3N - i^*(\alpha), 
\label{revised}
\end{equation}
where $N_4 = i^{*}(\alpha)-1$ is the number of quartic modes in region
1 of the frequency spectrum.  Thus, by measuring the number of quartic
modes, we are able to determine $\langle z \rangle (\alpha)$ as shown
in Fig.~\ref{fig:plot4}.  In the inset we show that $N/i^{*}(\alpha)
-N/(N+1)$ has two power-law regimes: scaling as $\sqrt{\alpha-1}$
($\alpha-1$) for small (large) $\alpha-1$.  In these limits, we obtain
\begin{equation}
\langle z \rangle (\alpha) = \langle z \rangle (1) + 2
\frac{A_n (\alpha -1)^n}{1 + A_n (\alpha-1)^n},
\label{eq:Model_Kfit}
\end{equation} 
where $n=0.5$ ($1$) for small (large) $\alpha-1$ and $A_{0.5}$ and
$A_1$ are positive constants.  In Fig.~\ref{fig:plot4}, we use a
function that interpolates between these power-laws and allows us to
fit $\langle z \rangle$ for the compression method over the entire range
of $\alpha$.  These arguments imply that $\langle z \rangle = \langle
z \rangle_{\rm iso}$ in the $\alpha \rightarrow \infty$ limit, yet
this is still an open question.  Previous studies have predicted that
$\langle z \rangle(\alpha) - \langle z \rangle (1)$ scale as
$\sqrt{\alpha-1}$ based on the behavior of the pair distribution
function $g(r)$ near contact for spherical particle packings
\cite{donev1}.  In contrast, our numerical results demonstrate
hypostaticity in ellipse packings originates from the existence of
quartic modes.

\paragraph{Annealed packings} 
The ellipse packings discussed up to this point were generated using
the compression method at fixed aspect ratio. Using this method, we
obtained similar $\langle z \rangle(\alpha)$ to that obtained
previously for hard ellipse packings \cite{donev1}.  Since ellipse
packings are hypostatic, it is in principle possible to obtain
packings with higher $\langle z \rangle$ than found in
Fig.~\ref{fig:plot4} without increasing the translational or
orientational order. To test this, we developed an `annealing' method,
which creates static packings by incrementally increasing the aspect
ratio from $\alpha=1$.  We initially create bidisperse disk
packings. Each disk is then assigned the same aspect ratio $1+\Delta
\alpha$ with the direction of the long axis chosen randomly. A new
ellipse packing is formed from this initial state using the
compression method described above. The ellipses of the new packing
are elongated again along their defined major axes, and the protocol
is repeated until a packing with the desired aspect ratio is reached.
Using this method, ellipse packings can be generated that are denser
and possess contact numbers much closer to $z_{\rm iso}=6$ as shown in
Fig.~\ref{fig:plot4}.  The annealed packings still exhibit a
low-frequency gap, and the location, $i^{*}$, of this gap can be used
to predict $\langle z \rangle$ through Eq.~\ref{revised}.  The
variation of $i^{*}$ with aspect ratio, however, differs from the
`compressed' packings.  The annealed packings tend to have higher
$\langle z \rangle$, fewer quartic modes, and a plateau in $\langle z
\rangle$ at large $\alpha$.

In conclusion, we find that aspherical grains qualitatively change the
nature of jamming at point J.  Model systems with spherical
grains appear to be the exceptional case: they are isostatic, and all
nontrivial modes of excitation increase quadratically with deformation
amplitude.  In contrast, ellipse packings, which are more relevant for
real granular media, possess quartic modes characterized by collective
rotational motions. Thus, Landau theories for ellipse packings have a
4th order term as the lowest nonvanishing contribution to the
generalized free energy.  These results will likely stimulate further
work on statistical descriptions of packings of aspherical particles.

\begin{acknowledgments}
Support from NSF grant numbers DMR-0549762 (BC $\&$ MM), DMR-0448838
(CSO) and CDI-0835742 (CS) is acknowledged.  During the workshop
``Dynamical heterogeneities in glasses, colloids, and granular media''
in Leiden in August 2008, we learned of independent, parallel results
by Zeravcic, et al. \cite{zeravcic} on 3D ellipsoids, with similar
conclusions.  We acknowledge discussions and suggestions from Z.
Zeravcic and S. Nagel at the workshop, and also thank M. Bi, G.
Lois, T. Witten, and N. Xu for helpful comments.
\end{acknowledgments}


\begin{references}
\bibitem{foot}
The parameters $\sigma_0$, $\beta$, and $\chi$ in Eq.~\ref{eq:sigma2} are 
defined as $\sigma_{0}=\sqrt{\frac{b_{i}^{2}}
{\lambda}+\frac{b_{j}^{2}}{1-\lambda}}$, $\beta=\left(\frac{
\left(a_{i}^{2}-b_{i}^{2}\right)
\left(a_{j}^{2}+\frac{1-\lambda}{\lambda} b_{i}^{2}\right)}{
\left(a_{j}^{2}-b_{j}^{2}\right)
\left(a_{i}^{2}+\frac{\lambda}{1-\lambda}
b_{j}^{2}\right)}\right)^{1/4}$, and $\chi=\left(\frac{
\left(a_{i}^{2}-b_{i}^{2}\right) \left(a_{j}^{2}-b_{j}^{2}\right)}{
\left(a_{j}^{2}+\frac{1-\lambda}{\lambda} b_{i}^{2}\right)
\left(a_{i}^{2}+\frac{\lambda}{1-\lambda}
b_{j}^{2}\right)}\right)^{1/2}$.
\bibitem{foot2} In the following discussion, we have accounted for
rattler ellipses and set $N=N'$.
\end{references}
\end{document}